\documentclass[aps,prl, twocolumn,showpacs,reprint]{revtex4-1}

\usepackage{graphicx}
\usepackage{hyperref}
\usepackage{amsmath}
\usepackage{amssymb}
\usepackage{epstopdf}

\begin{document}

\newcommand{\ket}[1]{$\left|#1\right\rangle$}

\title{Demonstration of a Memory for Tightly Guided Light in an Optical Nanofiber}

\author{B. Gouraud}
\affiliation{Laboratoire Kastler Brossel, UPMC-Sorbonne Universit\'es, CNRS, ENS-PSL Research University, Coll\`ege de France, 4 place
Jussieu, 75005 Paris, France}
\author{D. Maxein}
\altaffiliation[]{Present address: Deutsches Zentrum f\"ur Luft-und Raumfahrt, Space Administration, K\"onigswinterer Str. 522-524, D-53227 Bonn, Germany.}
\affiliation{Laboratoire Kastler Brossel, UPMC-Sorbonne Universit\'es, CNRS, ENS-PSL Research University, Coll\`ege de France, 4 place
Jussieu, 75005 Paris, France}
\author{A. Nicolas}
\affiliation{Laboratoire Kastler Brossel, UPMC-Sorbonne Universit\'es, CNRS, ENS-PSL Research University, Coll\`ege de France, 4 place
Jussieu, 75005 Paris, France}
\author{O. Morin}
\altaffiliation[]{Present address: Max-Planck-Institut f\"ur Quantenoptik, Hans-Kopfermann-Str. 1, D-85748 Garching, Germany.}
\affiliation{Laboratoire Kastler Brossel, UPMC-Sorbonne Universit\'es, CNRS, ENS-PSL Research University, Coll\`ege de France, 4 place
Jussieu, 75005 Paris, France}
\author{J. Laurat}
\email{julien.laurat@upmc.fr}
\affiliation{Laboratoire Kastler Brossel, UPMC-Sorbonne Universit\'es, CNRS, ENS-PSL Research University, Coll\`ege de France, 4 place
Jussieu, 75005 Paris, France}

\date{\today}

\begin{abstract}
We report the experimental observation of slow-light and coherent storage in a setting where light is tightly confined in the transverse directions. By interfacing a tapered optical nanofiber with a cold atomic ensemble, electromagnetically induced transparency is observed and light pulses at the single-photon level are stored in and retrieved from the atomic medium with an overall efficiency of $(10\pm0.5)\%$. Collapses and revivals can be additionally controlled by an applied magnetic field. Our results based on subdiffraction-limited optical mode interacting with atoms via the strong evanescent field demonstrate an alternative to free-space focusing and a novel capability for information storage in an all-fibered quantum network.
\end{abstract}

\pacs{03.67.-a, 03.67.Hk, 42.50.Gy, 42.50.Ex, 42.81.Qb}
\maketitle

Over the recent years, the physical implementation of quantum interfaces between light and matter has triggered a very active research, with unique applications to quantum optics and quantum information networks \cite{Kimble,Chang2014}. Within this context, a promising approach consists in coupling light with atomic ensembles \cite{DLCZ, Hammerer2010}. Reversible quantum memories have been realized in a variety of ensemble-based systems, e.g. doped crystals and free-space collection of alkali atoms \cite{Lvovsky}. Significant advances have been made, including the demonstration of entanglement between remote memories and the development of first rudimentary capabilities for quantum repeater architectures \cite{Chou05,Chou07,BDCZ,Sangouard2011}. However, free-space focusing as used in these seminal works is limiting the coupling one can obtain and the connectivity to fiber networks.

Interfacing guided light with atoms has therefore been foreseen as a promising alternative, enabling longer interaction length, large optical depth and potential non-linear interactions at low power level \cite{Chang2014}. A first possible implementation consists in encasing a vapor into the hollow core of a photonic-crystal fiber, confining thus atoms and photons in the waveguide. Slow-light, all-optical switching and few-photon modulation have been demonstrated \cite{Ghosh06,Bajcsy2009,Venkataraman2011}. Recently, single-photon-level Raman memory has been realized with larger core fibers, with storage limited to the 10 ns time-scale \cite{Sprague2014}. Another approach can be based on an even tighter confinement of light in a nanoscale waveguide leading to a large evanescent field that can interact with atoms located in the vicinity. This situation can be ideally realized with optical nanofibers exhibiting subwavelength diameter \cite{Tong2003}. Using a nanofiber in a hot Rubidium vapor, nonlinear interactions and low-power saturation have been reported \cite{Spillane2008, Hendrickson2010,Jones2014}, albeit with very short transit time of hot atoms in the evanescent field and large broadening.

In this new avenue of research, the unique prospects of combining cold atoms with nanofibers have triggered vast theoretical and experimental efforts. Pioneering works investigated the interaction of a small number of atoms with the guided mode, including fluorescence coupling and surface interactions \cite{Nayak2007, Sague2007,Morrissey2009}, and the dipole trapping of atoms in the evanescent field \cite{Vetsch2010,Goban2012}. Recent works have focused on the study of anisotropy in the scattering of light into the guided mode \cite{Mitsch2014, Petersen2014} and demonstrated the possibility of chiral nanophotonics based on this promising platform for light-mater interfacing within a fiber network.

In this Letter, we report the demonstration of an optical memory based on the interaction of cold cesium atoms with the evanescent field surrounding an optical nanofiber. By using electromagnetically induced transparency (EIT) and realizing the configuration initially proposed in 2002 by Hakuta and coworkers \cite{Patnaik2002}, slow-light and reversible storage at the single-photon level are demonstrated. With an additional magnetic field, controlled collapses and revivals are obtained. More generally, this work provides the first realization of a memory based on EIT in evanescent fields. In this configuration, we identify and quantitatively characterize the underlying decoherence mechanisms.

\begin{figure}[t!]
\includegraphics[width=0.85\columnwidth]{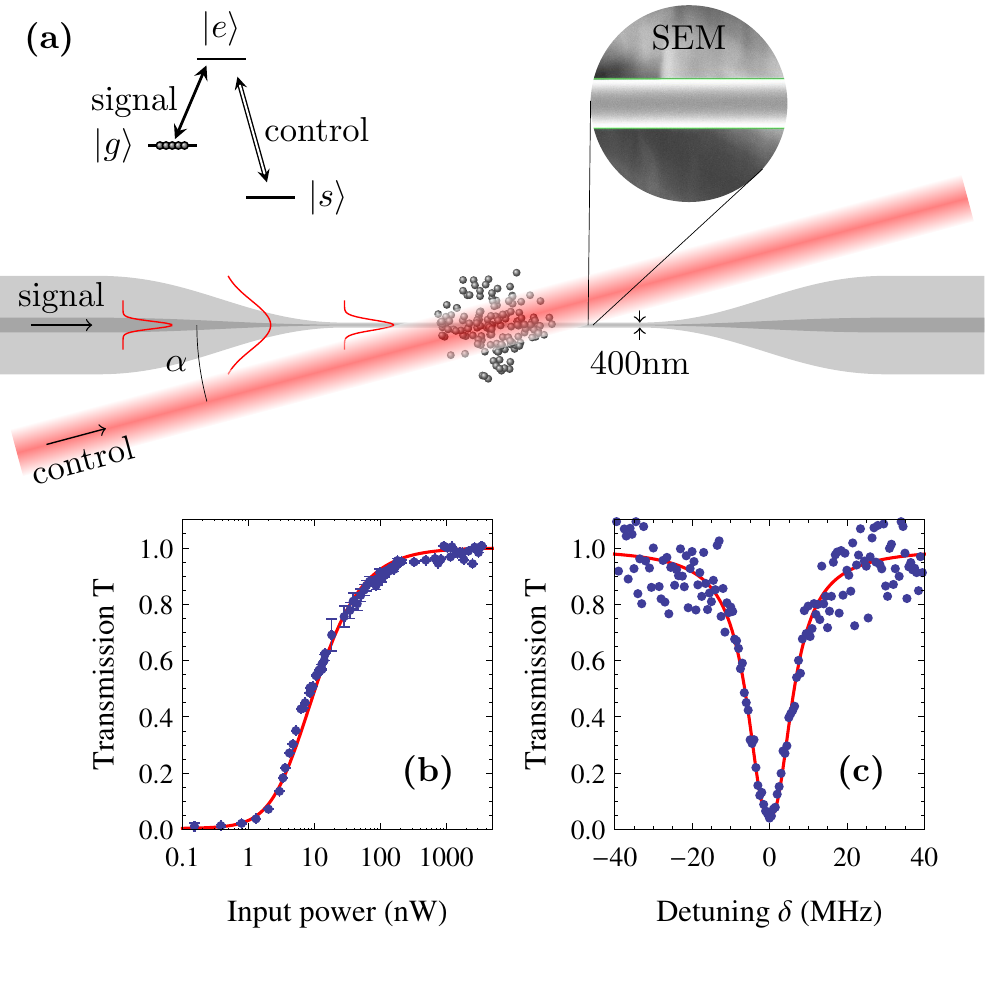}
\caption{(color online). Experimental setup. (a) A 400-nm diameter nanofiber is overlapped with an ensemble of cold cesium atoms. The signal to be stored is guided inside the nanofiber while the control propagates at the outside, with an angle $\alpha\sim13^{\circ}$.  (b) Transmission through the fiber of a pulse resonant on the $|g\rangle=\{{6S_{1/2}},{F=4}\} \rightarrow \{{6P_{3/2}},{F=5}\}$ transition as a function of the input power. The low saturation power results from the confined geometry over the cloud length. (c) Transmission profile for a signal pulse at the single-photon level as a function of the detuning $\delta$ from the $|g\rangle \rightarrow |e\rangle=\{{6P_{3/2}},{F=4}\}$ transition.}
\label{fig1}
\end{figure}

Our setup is illustrated in Fig.~\ref{fig1}(a). A cloud of laser-cooled cesium atoms overlaps with a nanofiber suspended in the ultra-high vacuum chamber and connected to the outside by two teflon feedthroughs. The nanofiber is fabricated from a non polarization-maintaining fiber (Thorlabs SM800-5.6-125) by flame brushing \cite{Birks1992,Orucevic2007,Stiebeiner2010,Aoki2010,Hoffman2014}. It exhibits a $2r=400\pm20$ nm diameter over a length of $9$ mm, longer than the cesium cloud. The silica-vacuum interface guides the hybrid fundamental mode HE$_{11}$ along the nanofiber \cite{Tong2004}, which is adiabatically coupled via a tapered region on both sides. 
The fraction of the energy travelling in the evanescent part can get arbitrarily large when the fiber diameter decreases. However the energy will also spread further from the fiber \cite{Spillane2008}. As a result, the maximal intensity in the evanescent field is obtained for a diameter around 400 nm and the evanescent fraction of the energy reaches 40\% \cite{Tong2004}. Due to this evanescent field, the guided mode is coupled to atoms in the vicinity.  Strong absorption -- not limited to the evanescent fraction -- can be obtained if the number of atoms is large enough.  Theoretical studies have also shown that the guided mode can propagate under EIT condition imposed by the surrounding medium and, in the ideal case, full storage can be obtained \cite{LeKien2009,LeKien2015}.

The cesium cloud is released from a magneto-optical trap (MOT) cigar-shaped along the fiber direction by using rectangular coils. The experiment is conducted in a cyclic fashion. First, the current in the MOT coils is switched off, then, after a $4.4$ ms decay of eddy currents, the trapping beams are turned off. Measurements are performed during $1$ ms, while the cloud expands freely.  After this stage, the trap is reloaded for $40$ ms. The surrounding magnetic fields are measured by Zeeman sublevel microwave spectroscopy and cancelled with three pairs of coils. Compensation below 20 mG is obtained, corresponding to a broadening of 100 kHz. The $\Lambda$-system for EIT involves the two ground states $|g\rangle=\{{6S_{1/2}},{F=4}\}$ and $|s\rangle=\{{6S_{1/2}},{F=3}\}$, and one excited state $|e\rangle=\{{6P_{3/2}},{F=4}\}$. The atoms are initially prepared in $|g\rangle$.

The relative polarisation of the signal and control beams must be chosen properly. For the Cs levels used here, without optical pumping, EIT is efficient for orthogonal polarisations \cite{Nicolas2014}. However, the guided light has a complex polarisation pattern, including a significant non-transverse component \cite{Vetsch2012}. Controlling this polarization is thus crucial. This can be done by using the method described in \cite{Vetsch2012,Goban2012} based on Rayleigh scattering. When a 0.5 mW laser beam is sent through the fiber, the light partly scattered by inhomogeneities and surface impurities can indeed be detected from the side by a camera equipped with a polarization filter to suppress the non-transverse component. By adjusting the polarization at the input, it is therefore possible to obtain a polarization pattern in the transverse plane with a quasi-linear orientation over the nanofiber waist. This polarization is aligned here horizontally and stable over many hours.

As a preliminary characterization, we monitor the absorption of a light pulse propagating through the fiber. This measurement is first used to optimise the overlap of the cloud with the nanofiber waist: the cloud position is adjusted for maximal absorption by slightly misaligning the trapping beams.

Transmission for a probe at resonance with the cycling $|g\rangle \rightarrow \{{6P_{3/2}},{F=5}\}$ transition is reported on Fig.~\ref{fig1}(b) as a function of its power. This measurement provides the saturation power, which is expected to be very low due to the tight confinement. The shape is fitted accordingly to the empirical nonlinear model $T=e^{-\alpha L}$ with $\alpha=\alpha_0/(1+P/P_{\textrm{sat}})^k$, which has been shown to be well adapted to the nanofiber case with $k=1$ \cite{Jones2014}. This fit yields a saturation power $P_{\textrm{sat}}=1.3\pm 0.2$~nW and an absorbed power in the saturated regime $P_{\textrm{abs}}=\alpha_0 L P_{\textrm{sat}}=8\pm2$~nW.  $P_{\textrm{abs}}$ enables to estimate an effective number $N$ of atoms. By considering the nominal power radiated by a saturated single Cs atom $p=3.8$ pW, $N$ can be inferred as $N=P_{\textrm{abs}}/p=2000\pm 500$ atoms. This value is compatible with an independent estimate taking into account the spatial dependence of the atomic density in the vicinity of the fiber \cite{Sague2007,LeKien2009}. Indeed, due to van der Walls interaction, the density is reduced close to the surface. The density radial dependence can be explicitly calculated, as given in \cite{LeKien2009}. By considering an interaction length $L=5$ mm, a typical MOT density of $10^{11}$ atoms/cm$^3$ and by integrating over a distance from the surface equal to 4$r$, this estimation leads to 1500 atoms. Due to the geometry, this small number of atoms compared to free-space implementations can lead however to an optically dense ensemble.

Figure \ref{fig1}(c) then shows the transmission profile for a signal pulse as a function of the detuning $\delta$ from the $|g\rangle \rightarrow |e\rangle$ transition. The fitted profile, $\exp\left[-\textrm{OD}/(1+(2\delta/\Gamma)^2)\right]$, yields an optical depth $\textrm{OD}~=~3$ and $\Gamma/2\pi=6.8\pm 0.5$~MHz. This value is 30\% larger than the natural linewidth in free space, $\Gamma_0/2\pi=5.2$~MHz, resulting from the finite temperature, surface interactions and modification of the spontaneous emission rate in the vicinity of the fiber \cite{Sague2007}.

We now turn to the study of EIT, where a control field can change the transmission characteristics of the probe \cite{Fleischhauer}. A large control beam propagating in free space is shined on the cloud, with a $400$ $\mu$m waist and an angle $\alpha\sim13^\circ$ with the nanofiber. This angle has been minimized given the technical constraints in the apparatus. The control is produced by an extended-cavity laser diode and is frequency-locked at the 9.2 GHz hyperfine frequency with the signal generated from a Ti:Sa laser. 

The measurements are performed at the single-photon level. Reaching this regime requires to filter out the contamination from the control that couples from freespace into the guided mode. This coupling is on the order of $10^{-8}$. For this purpose, we use polarization filtering at the fiber output, taking advantage of the experimental fact that the control couples into the nanofiber with a quasi-linear vertical polarisation, thus orthogonal to the signal polarisation. Additional frequency filtering is obtained from a Cs cell pumped in \ket{s}. With these filterings, the remaining light at the detection is $10^{14}$ times smaller than the initial control power. Measurements are done at the fiber output using an avalanche photodiode (SPCM-AQR-14-FC).

\begin{figure}[b!]
\includegraphics[width=0.87\columnwidth]{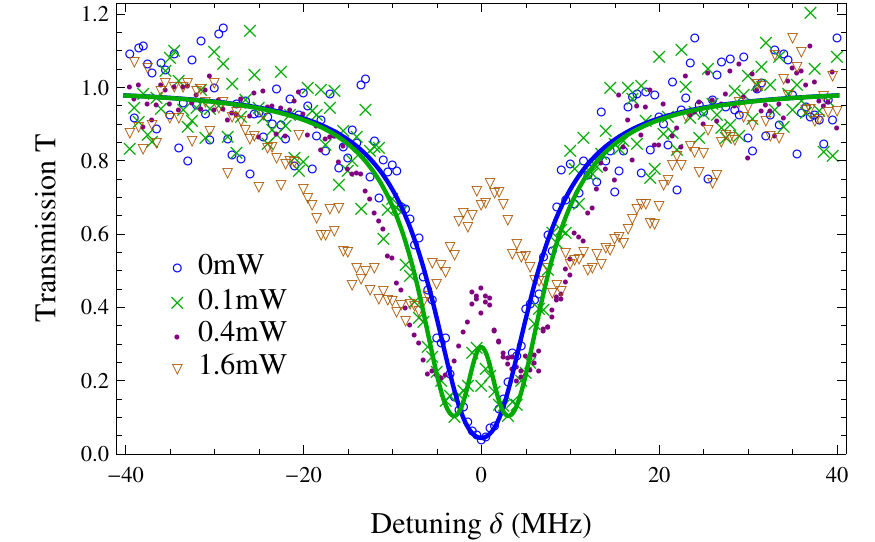}
\vspace{-0.1cm}\caption{(color online). Electromagnetically-induced transparency for the guided light. The control is on resonance on the $|s\rangle \rightarrow |e\rangle$ transition while the signal is detuned by $\delta$ from the $|g\rangle \rightarrow |e\rangle$ resonance. Profiles are displayed as a function of $\delta$, for four values of the control power. }
\label{fig2}
\end{figure}

Figure \ref{fig2} gives the transmission profiles of the signal as a function of its detuning $\delta$ from resonance, for different values of the control power. When the control field is applied, a transparency window appears, providing a first signature of EIT in this evanescent-field configuration. Transparency close to 75\% is measured for a control power of 1.6 mW. The finite contrast is primarily due to a non-negligible ground-state decoherence, which arises from concurrent mechanisms that will be detailed later. We note that a full model should include the complexity of the guided light polarization and its evanescent nature, as first done in \cite{LeKien2009}, but also importantly the complex level structure of atomic Cs, including Zeeman levels and other excited levels of the $6P_{3/2}$ manifold, as testified by the damping and asymmetries of the resonances \cite{Giner2013}.

\begin{figure}[b!]
\hspace{-0.4cm}\includegraphics[width=0.95\columnwidth]{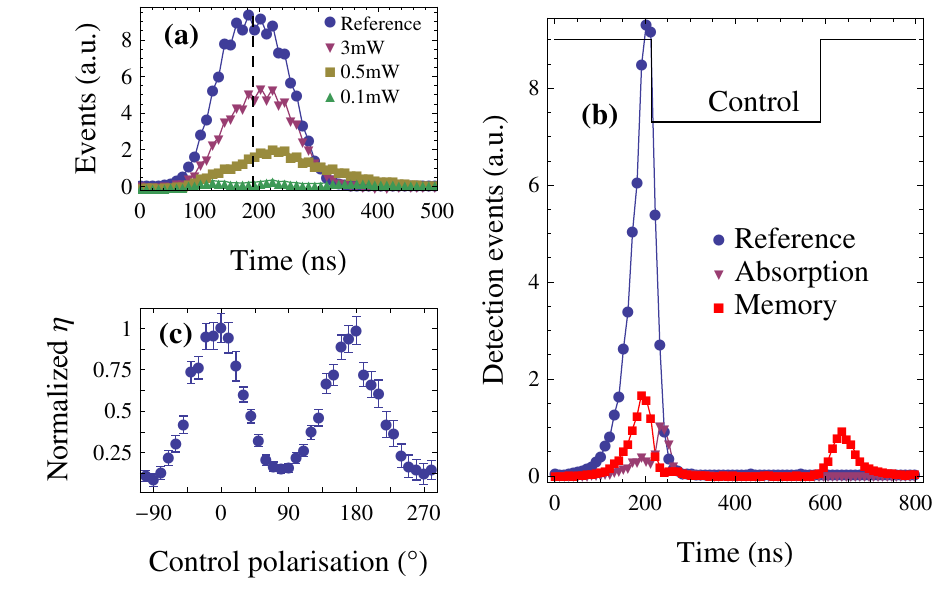}
\vspace{-0.3cm}\caption{(color online). Slow-light and storage at the single-photon level. (a) Transmitted pulses for different control powers. The reference is measured without atoms. (b) Storage and retrieval. In the absence of control, the blue and purple points give the transmitted pulse without and with atoms. The red data corresponds to the memory sequence, showing leakage and retrieval. The black line indicates the control timing. After the end of the input pulse, the reference and absorption curves are superimposed and correspond to the background noise level. The mean photon-number per pulse is $0.6$ and the efficiency is $10\pm0.5$ $\%$. (c) Efficiency as a function of the control linear polarisation angle. The zero angle corresponds to a vertical polarisation.
}
\label{fig3}
\end{figure}

After having observed transparency, we measure the delay, i.e. slow-light effect \cite{Hau1999}, resulting from pulse propagation under EIT condition. As a signal, we use weak laser pulses at the single-photon level. Results are displayed in Fig.~\ref{fig3}(a). When the control power is decreased, smaller transparency but larger delays are obtained due to the narrower transparency window. For a 0.5 mW control, close to the value used in the subsequent experiments, we achieve a 60~ns delay. This value corresponds to a 3000-fold reduction in group velocity.

Next, we demonstrate the storage of the guided light by the dynamic EIT protocol \cite{Liu2001}. While the light is slowed down, the control is ramped down to zero and the signal is converted into a collective excitation. Later, the control is switched on again and the light can be retrieved back in a well-defined spatio-temporal mode due to the collective enhancement provided by the ensemble. Figure \ref{fig3}(b) provides the storage results for a signal with a mean photon-number per pulse equal to 0.6$\pm$0.1. The pulses have been temporally shaped \cite{Novikova} to an exponentially-rising profile with a full width at half-maximum of 60~ns. Due to the limited delay, the pulse cannot be contained entirely in the ensemble, and a leakage is observed before the control is switched off. Efficiency of storage and retrieval is defined as the ratio of the photodetection events in the read-out to the ones in the reference. After optimisation of the control power for a trade-off between transparency and delay, an efficiency $\eta=10\pm 0.5$~$\%$ is obtained, therefore realizing a memory at the single-photon level in this fibered setting. The normalized efficiency as a function of the control polarization is provided in Fig. \ref{fig3}(c). This result shows the good control over the polarization in the nanofiber waist. We also note that the achieved efficiency is compatible with the limited OD used here. Remarkably, the single-photon signal-to-noise ratio in the retrieved pulse is already equal to 20.

\begin{figure}[t!]
\hspace{-0.4cm}\includegraphics[width=0.95\columnwidth]{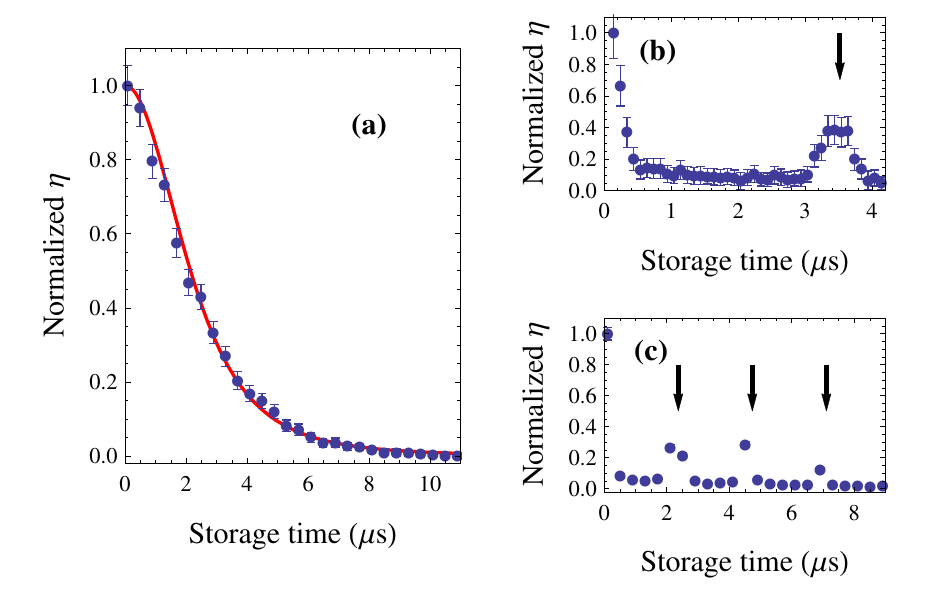}
\vspace{-0.3cm}\caption{(color online). Memory lifetime and revivals. (a) Normalized retrieval efficiency $\eta$ as a function of the storage time. The data are fitted according to the expression given in the text. (b) and (c) Retrieval efficiency in the presence of a DC magnetic field of $0.4$ and $0.6$ Gauss respectively. Revivals are observed at multiples of the half Larmor period. The arrows indicate the theoretical positions calculated from the calibrated field.
}
\label{fig4}
\end{figure}

We finally investigate the memory lifetime. Figure \ref{fig4}(a) gives the retrieval efficiency as a function of the storage duration. Three concurrent decoherence mechanisms are involved and can be evaluated independently. The atomic motion related to the finite temperature first results in a possible loss of the atoms from the tiny evanescent field area. By denoting $v=\sqrt{k_B T/m}$ the thermal velocity of an atom of mass $m$ at a temperature $T$,  the transit time can be estimated by $\tau_1=2r/v$. For a 200~$\mu$K temperature, this expression provides $\tau_1=$~3.6~$\mu$s. 

The two other decoherence contributions come from dephasing of the stored collective excitation, which can be written in the ideal case as $|\psi\rangle=(1/\sqrt{N})\sum_j e^{i\phi_j} |g...s_j...g\rangle$ where $N$ is the number of atoms. The first process, also related to the temperature, is the motional dephasing due to the strong angular dependence of EIT \cite{Tabosa2004,Peters2012}. 
 The Doppler shift of the two-photon transition results in a phase change between the storage and retrieval times $\Delta\phi_j=\Delta\vec{k}.(\vec{r}_s-\vec{r}_r)$ where $\vec{r}_s$ and $\vec{r}_r$ are the initial and final positions of the jth atom, and $\Delta\vec{k}$ is the wave-vector mismatch of the control and guided signal \cite{Pan}. It leads to a lifetime $\tau_2$ with $1/\tau_2\simeq(4\pi/\lambda)\sin(\alpha/2) \, v$. With $\alpha\sim13^{\circ}$, one finds $\tau_2=$~5.3 $\mu$s. The second process is caused by the residual magnetic field that results in an atom-dependent Larmor precession. The measured Zeeman inhomogeneous broadening is 100 kHz and the associated lifetime can be estimated to $\tau_3=$~10~$\mu$s \cite{Kyung}. The combination of the loss of the atoms on one hand and dephasing of the collective state on the other gives a decay of the efficiency of the form $\exp\left[-(t/\tau_{D})^2/(1+(t/\tau_T)^2)\right]/\left(1+(t/\tau_T)^2\right)^2$ \cite{Jenkins}, with $\tau_T=\tau_1$ and $1/\tau_D^2=1/\tau_2^2+1/\tau_3^2$. The fit in Fig. \ref{fig4}(a) yields $\tau_D=5.5\pm1$~$\mu$s and $\tau_T=3.7\pm0.2$~$\mu$s. These values are in good agreement with the evaluated time scales and thereby confirm that the decoherence mechanisms are well identified.   

Applying an additional DC magnetic field enables to control the time evolution of the stored excitation and to demonstrate collapses and revivals in the retrieval efficiency as a function of the storage time, as observed in free-space implementations \cite{Matsukevich2006}. Due to different Zeeman sublevels, the phases $\phi_j$ rotate at multiples of the Larmor frequency and revivals occur when the terms rephase. Figure \ref{fig4}(b) and \ref{fig4}(c) correspond to a uniform magnetic field aligned along the nanofiber. The field is calibrated by measuring the Zeeman shifts. Revivals are observed at multiples of the half Larmor period, equal to 3.5 $\mu$s for 0.4 G and 2.35~$\mu$s for 0.6 G.

In conclusion, we have reported the realization of EIT and storage of light at the single-photon level in a nanofiber-based interface. This capability based on the interaction of the evanescent part of the tightly guided mode with the surrounding atoms provides an intrinsically-fibered memory, with potential applications to multiplexed schemes \cite{Collins2007}.  The results, which are obtained with low optical depth relative to previous free-focusing and hollow-core fiber memory demonstrations, are promising given the possible improvements. Larger optical depth, retrieval efficiency and coherence time are expected by combining the present protocol with the recently achieved dipole trapping of atom arrays close to the surface \cite{Vetsch2010,Goban2012}. Copropagating the control field in the guided mode would also enable a better matching of the involved light polarizations, a reduced decoherence rate and a control field at an ultra-low power level. Efficient frequency filtering will be required to distinguish between the copropagating trapping light at the milliwatt level and the single-photon-level pulses. Finally, the present demonstration opens up the possibility to implement the seminal Duan-Lukin-Cirac-Zoller protocol \cite{DLCZ} in this setting, with the unique prospects of efficient generation of intrinsically fibered single photons and the remote entanglement of all-fibered ensembles.\\

\begin{acknowledgments}
This work is supported by the European Research Council (Starting Grant HybridNet), the Emergence program from Ville de Paris and the IFRAF DIM Nano-K from R\'egion Ile-de-France. The authors also acknowledge interesting discussions within the CAPES-COFECUB project Ph 740-12. A.N. acknowledges support from the Direction G\'en\'erale de l'Armement (DGA). J.L. is a member of the Institut Universitaire de France.
\end{acknowledgments}

\end{document}